\documentclass[aps,pre,twocolumn,superscriptaddress,twoside, nofootinbib]{revtex4-1}
\usepackage{amsmath,amssymb,amsfonts}
\usepackage{graphicx}
\usepackage[english]{babel}
\usepackage{psfrag}
\usepackage{color}
\usepackage{tabularx}
\usepackage{times}
\usepackage{hyperref}
\usepackage{float}
\usepackage{CJK}

\begin{document}
\title{Comment on \\ "Gapless spin liquid ground state of the spin-$\frac{1}{2}$ $J_1$-$J_2$ Heisenberg model on square lattices"}

\author{Bowen Zhao} 
\email{bwzhao@bu.edu} 
\affiliation{Department of Physics, Boston University, 590 Commonwealth Avenue, Boston, Massachusetts 02215, USA}

\author{Jun Takahashi} 
\affiliation{Department of Physics, Boston University, 590 Commonwealth Avenue, Boston, Massachusetts 02215, USA}
\affiliation{Beijing National Laboratory for Condensed Matter Physics and Institute of Physics, Chinese Academy of Sciences, Beijing 100190, China}

\author{Anders W. Sandvik} 
\email{sandvik@bu.edu} 
\affiliation{Department of Physics, Boston University, 590 Commonwealth Avenue, Boston, Massachusetts 02215, USA}
\affiliation{Beijing National Laboratory for Condensed Matter Physics and Institute of Physics, Chinese Academy of Sciences, Beijing 100190, China}

\date{\today}
\begin{abstract}
Liu et al. [Phys.~Rev.~B {\bf 98}, 241109 (2018)] used Monte Carlo sampling of the physical degrees of freedom of a Projected Entangled Pair State (PEPS) 
type wave function for the $S=1/2$ frustrated $J_1$-$J_2$ Heisenberg model on the square lattice and found a non-magnetic state argued to be a gapless spin 
liquid when the coupling ratio $g=J_2/J_1$ is in the range $g \in [0.42,0.6]$. Here we show that their definition of the order parameter for another candidate 
ground state within this coupling window---a spontaneously dimerized state---is problematic. The order parameter as defined will not detect dimer order when 
lattice symmeties are broken due to open boundaries or asymmetries originating from the calculation itself. Thus, a dimerized phase for some range of $g$ 
cannot be excluded (and is likely based on several other recent works).
\end{abstract}

\maketitle
	
\section{Overview} 
		
In a recent Rapid Communication \cite{Liu18}, Liu et.~al.~argued that there is a gapless spin liquid phase in the ground state of the $S=1/2$ frustrated 
square-lattice $J_1$-$J_2$ Heisenberg model for $g=J_2/J_1  \in [0.42,0.6]$. At variance with other recent works \cite{Gong14,Morit15,Wang18,Haghshenas18}, 
they found no  spontaneously dimerized valence-bond-solid (VBS) phase within this range of coupling ratios (where other works have roughly placed the VBS
at $g \in [0.52-0.61]$). They reached their conclusions based on the method of Monte Carlo sampling of gradient-optimized tensor network states 
\cite{Sandvik07,Schuch08,Wang11}, which they have further refined for the specific case of a tensor network of the Projected Entangled Pair State (PEPS) 
type. Open-boundary lattices with up to $16\times 16$ spins were used, and, taken at face value, the results appear to be well converged and reliable.
			    
In this Comment we point out that the definition of the VBS order parameter used by Liu et al.~has a potential flaw and may not capture long-range order 
correctly on the open-boundary lattices considered. The squared columnar VBS order parameters for $x$ and $y$ oriented dimers, $m_{dx}^2$ and $m_{dy}^2$, 
were defined in Eq.~(2) of \cite{Liu18} as follows (up to typographical errors):
\begin{equation}
m_{d \alpha}^2 = \frac{1}{N_b^2} \sum_{\mathbf{rr'}} e^{i \mathbf{q}_\alpha \cdot (\mathbf{r} - \mathbf{r'})}
(\langle B^\alpha_\mathbf{r} B^\alpha_\mathbf{r'} \rangle - \langle B^\alpha_\mathbf{r}\rangle\langle B^\alpha_\mathbf{r'}\rangle), 
\label{vbsdef1}
\end{equation}
where $\alpha = x, y$, $B^\alpha_{\bf r}={\bf S}({\bf r}) \cdot {\bf S}({\bf r + \hat \alpha})$ is the bond operator along the $\alpha$ direction,
and $N_b$ is the number of bonds summed over. The wave-vector corresponding to columnar order is $\mathbf{q}_\alpha =  (\pi, 0)$ and $(0, \pi)$ for 
$\alpha = x$ and $\alpha = y$, respectively. We can rewrite this squared order parameter in the equivalent form:
\begin{equation}
m_{d \alpha}^2 = \langle D_\alpha^2 \rangle - \langle D_\alpha \rangle^2,
\label{vbsdef2}
\end{equation}
where
\begin{equation}
D_\alpha = \frac{1}{N_b} \sum_\mathbf{r} e^{i \mathbf{q}_\alpha \cdot \mathbf{r}} \mathbf{S}(\mathbf{r}) \cdot \mathbf{S}(\mathbf{r} + \hat{\alpha}).
\label{ddef}
\end{equation}
The problem with the definitions is the subtraction of the non-uniform $\langle B^\alpha_\mathbf{r}\rangle\langle B^\alpha_\mathbf{r'}\rangle$ in 
Eq.~(\ref{vbsdef1}) or $\langle D_\alpha \rangle^2$ in Eq.~(\ref{vbsdef2}) when long-range order is induced by some symmetry-breaking mechanism, e.g., 
with certain open lattice boundaries or some imperfection in the method used. In essence, the baby is then thrown out with the bath water.

We will demonstrate this problem by considering a columnar VBS state which is four-fold degenerate on periodic $L \times L$ lattices with even 
$L$. The ground state is uniform in the absence of some symmetry-breaking mechanism, and the subtracted term $\langle D_\alpha \rangle^2$ in 
Eq.~(\ref{vbsdef2}) vanishes. However, on rectangular lattices with $L_x \times L_y$ spins (even $L_x$ and $L_y$) the ground state is unique and hosts
a specific dimer pattern. The two terms then cancel each other in the limit of large system sizes, thus rendering the definitions Eq.~(\ref{vbsdef1})
and Eq.~(\ref{vbsdef2}) unsuitable for detecting the dimer order. On square $L \times L$ lattices there is a two-fold symmetry left, and the definitions 
can in principle detect the dimerization (albeit with a reduced value of the order parameter). However, in practice the calculation itself may break the 
$90^\circ$ lattice rotation symmetry, and then again the definition is not suitable. Due to the likely symmetry breaking of the PEPS calculations in 
Ref.~\cite{Liu18}, a VBS phase in the $J_1$-$J_2$ Heisenberg model cannot be ruled out based on the results presented.

In the following we will use a specific example of a quantum spin model with a well established columnar VBS phase to illustrate our arguments; the $S=1/2$
square-lattice $J$-$Q_3$ model \cite{Lou09,Sandvik12} 
defined by the Hamiltonian
\begin{equation}
H = - J \sum_{\langle jl \rangle} P_{ij} - Q_3  \hskip-2mm\sum_{{\langle ijklmn \rangle}} \hskip-2mm P_{ij} P_{kl} P_{mn}.
\label{jqham}
\end{equation}
Here $P_{ij}=1/4-{\bf S}_i \cdot {\bf S}_j$ is a singlet projector and the first term in Eq.~(\ref{jqham}) is the standard antiferromagnetic Heisenberg 
exchange between nearest neighbor spins. In the second term, the three index pairs ${ij}$, ${kl}$, and ${mn}$ correspond to parallel links forming 
columns  on $3\times 2$ and $2\times 3$ lattice cells. This correlated singlet interaction leads to the formation of a four-fold degenerate columnar 
VBS with a spontaneous $Z_4$ symmetry-breaking transition at a critical value of $Q_3/J$. Here we will consider the case $J=0$ and focus on the detection 
of the columnar order on open lattices, using a ground-state valence-bond projector quantum Monte Carlo (QMC) method \cite{Sandvik10} with which the 
spin-rotation invariant bond correlations in Eqs.~(\ref{vbsdef1}) and (\ref{vbsdef2}) can be evaluated easily. Though we use a different model and a 
different method for obtaining the ground state, the order parameter is the same as in Ref.~\cite{Liu18}, and the problem of subtracting off a boundary-induced 
expectation value when the symmetry is broken is exactly the same. In a previous work by one of us \cite{Sandvik12}, related issues were discussed 
in the context of cylindrical lattices (often used in DMRG calculations \cite{Gong14,Wang18}), i.e., ones with periodic boundary conditions in one direction 
and open boundaries in the other direction. Here we focus specifically on the problems with the order parameter definitions in Eqs.~(\ref{vbsdef1}) and 
(\ref{vbsdef2}) when all boundaries are open, as is more practical in PEPS calculations.

In Sec.~\ref{sec:rectangular} we first consider $L_x\times L_y$ lattices with $L_x=2L_y$, for which the VBS pattern is unique (for $L_y$ an even number) 
and our arguments can be illustrated most clearly. In Sec.~\ref{sec:square} we consider the slightly more subtle case of $L\times L$ lattices (again with $L$ 
even), on which the columnar VBS pattern is two-fold degenerate. We summarize our conclusions and discuss implications in Sec.~\ref{sec:conclusions}.

\begin{figure}[t]
\includegraphics[width=70mm,clip]{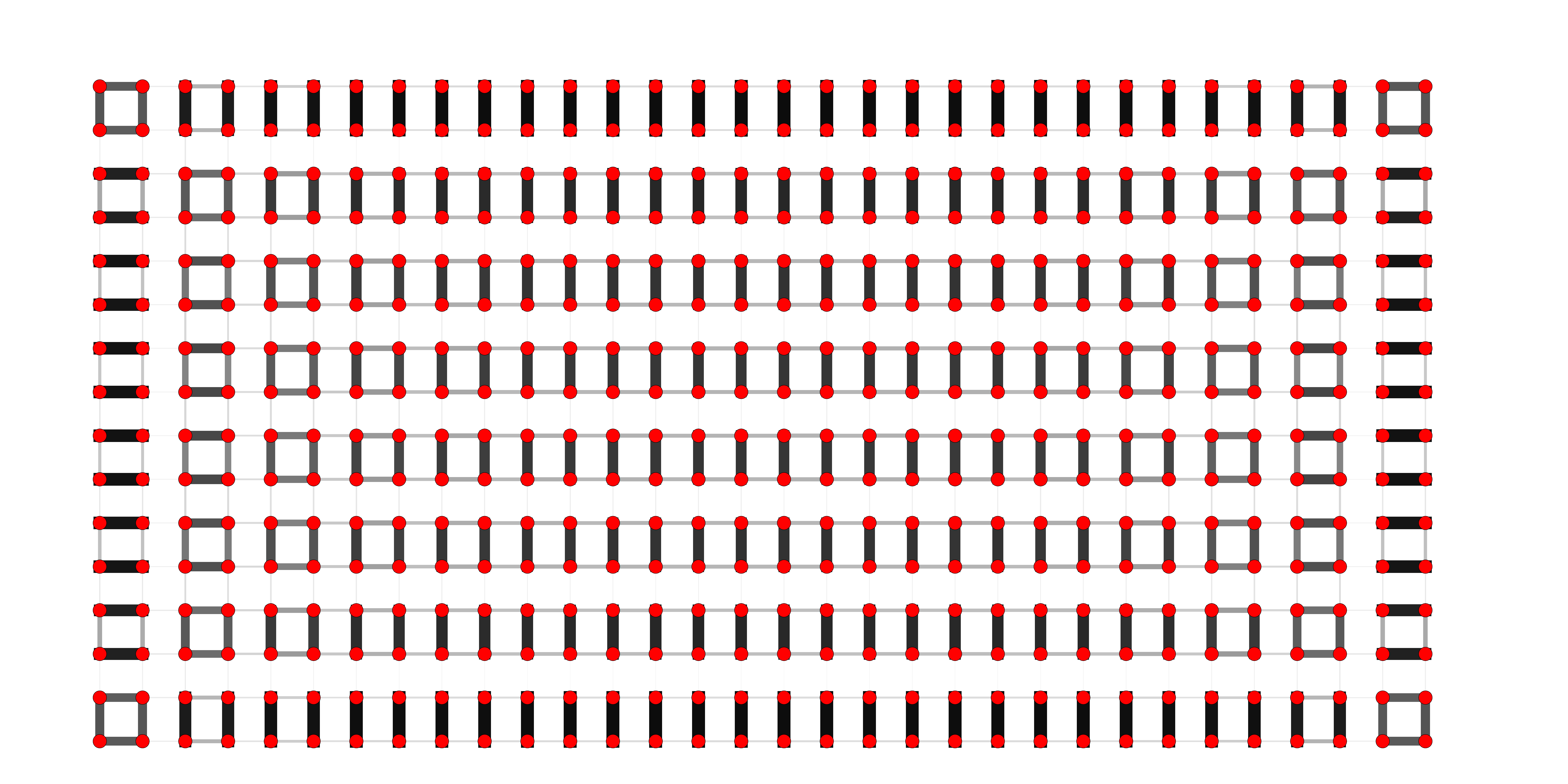}
\vskip-2mm
\caption{Bond strength illustration for a $32 \times 16$ system. The line thickness is proportional to $|\langle S^z_iS^z_j\rangle|=
-\langle{\bf S}_i \cdot {\bf S}_j\rangle/3$ on each nearest-neighbor link $ij$, obtained with QMC simulations of the $Q_3$ model
[$J=0$ in Eq.~(\ref{jqham})].}
\label{Fig:BondStrength}
\end{figure}

\section{Rectangular Lattices}
\label{sec:rectangular}

As mentioned above, on an open $2L\times L$ lattice with even $L$ there is a unique columnar VBS pattern in the ground state. In the case
of the $Q_3$ model, $J=0, Q_3=1$ in Eq.~(\ref{jqham}), the boundaries favor dimers (bonds with a higher singlet density) perpendicular 
to the edges, as illustrated for a $32\times 16$ system in Fig.~\ref{Fig:BondStrength}. The dimer orientation favored by the longer 
edge survives in the center of the system, and it is  possible to use the dimer order parameter $D_y$ defined in Eq.~(\ref{ddef}) without 
squaring (as noted previously, e.g., in Ref.~\cite{Sandvik02}). 

\begin{figure}[t]
\includegraphics[width=75mm,clip]{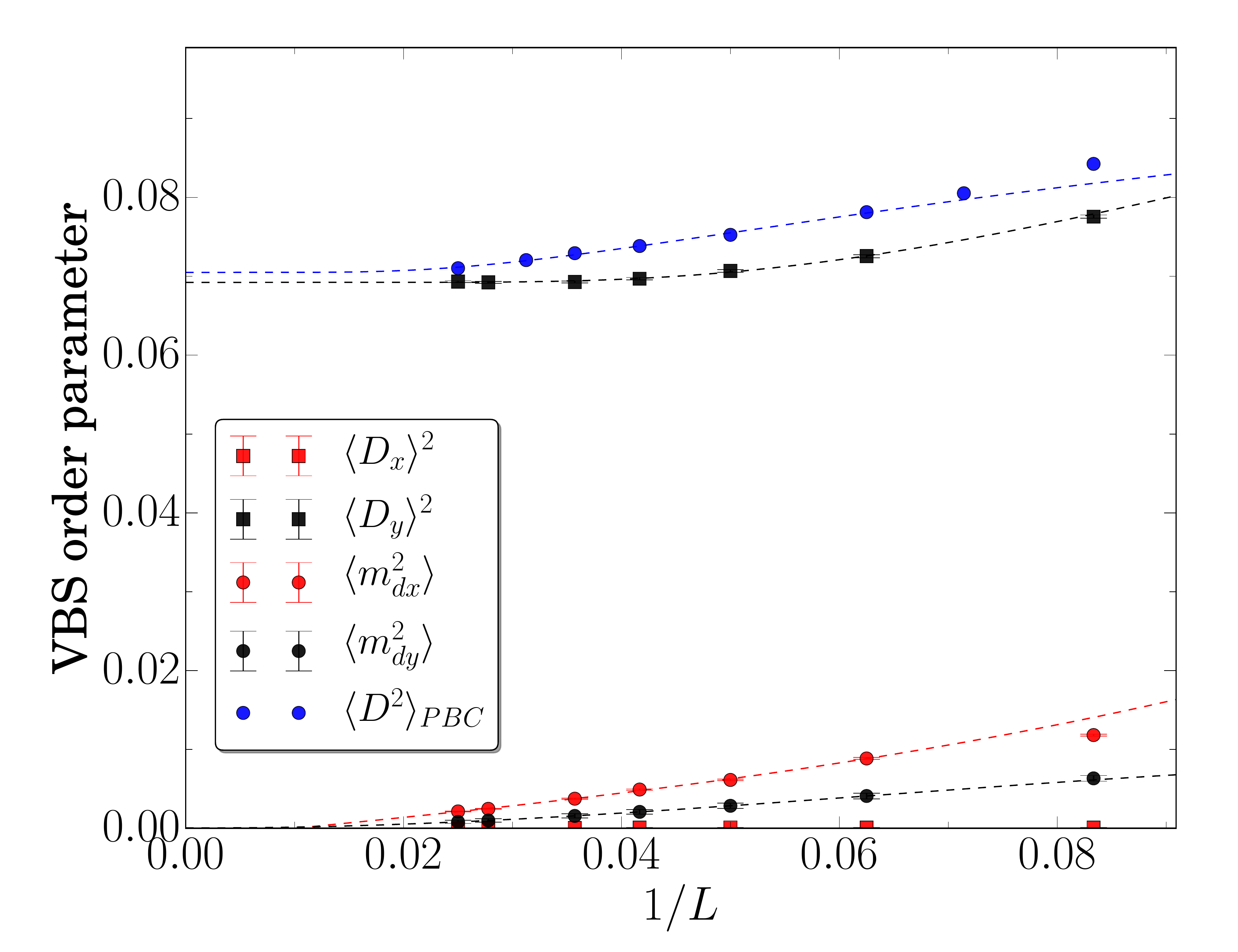}
\vskip-2mm
\caption{Inverse system size dependence of different definitions of the columnar VBS order parameter, computed by  QMC simulations on
the central $L\times L/2$ part of $2L\times L$ lattices with $L$ up to $40$. The curves are fits to the form $a + b{\rm e}^{-cL}$,
with adjustable parameters $a,b,c$. Note that $\langle D_x\rangle^2$ is very close to $0$.}
\label{Fig:D}
\end{figure}

To demonstrate that correct results are obtained with $D_y$ in the thermodynamic limit, in Fig.~\ref{Fig:D} we compare results for $\langle D_y\rangle^2$ 
computed on the central $L\times L/2$ bonds of $2L\times L$ lattices (to eliminate some of the boundary enhancements of the order, though this is
not necessary) with results for $\langle D^2\rangle = \langle D_x^2\rangle + \langle D_y^2\rangle$ calculated on periodic $L\times L$ lattices.
We have fitted both data sets using exponentially convergent forms, as expected for VBS order \cite{Sandvik12}, but details of the fits are not 
important here. The extrapolated, clearly non-zero values are compatible with each other. In contrast, using the definition $m_{dy}^2$ by Liu et al., 
Eq.~(\ref{vbsdef1}), gives results approaching zero with increasing system size (we only show rough fits, but the trend is
clear), as expected when the symmetry-broken order parameter has been subtracted off. The $x$-oriented order parameters should of course vanish, 
on account of the rectangular lattice shape inducing only $y$ columnar order in the thermodynamic limit.

\section{Square Lattices}
\label{sec:square}

\begin{figure}[t]
  \includegraphics[width=55mm,clip]{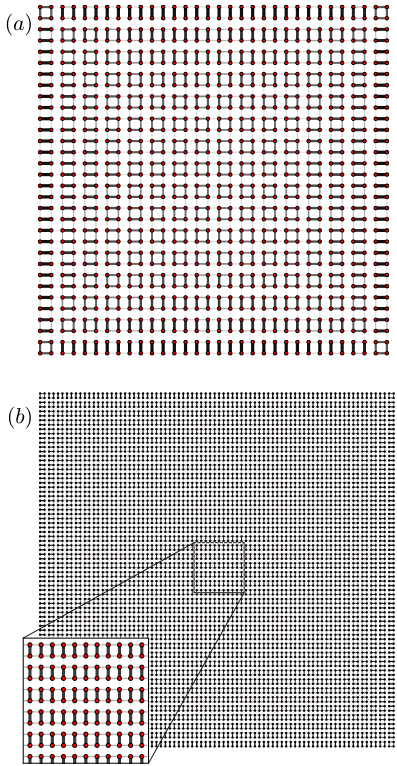}
  \vskip-2mm
  \caption{Bond strengths for (a) $32 \times 32$ and (b) $80 \times 80$ (b) systems. The plaquette pattern for the $L= 32$ case reflects
  averaging over $x$- and $y$ bond order. In the $L=80$ case, the simulation was not long enough   to sample equally the two degenerate 
  sectors, and an $y$-oriented pattern is apparent at the center of the system.}
  \label{Fig:BondStrength_SQ}
\end{figure}

As an example more closely corresponding to the calculations in Ref.~\cite{Liu18}, we next consider the same $Q_3$  model as above
but on $L \times L$ lattices (even $L$), again with all open boundaries. Since now there is no anisotropy between the $x$ and $y$ directions, the true ground 
state does not have a unique locked-in dimer pattern, but is two-fold degenerate with fluctuations between $x$- and $y$-oriented order. On a small lattice,
these fluctuations are fully sampled in our QMC simulations running for reasonable times, and when averaged the bond patterns look more like a
plaquette VBS state. This is shown in Fig.~\ref{Fig:BondStrength_SQ}(a) for a $32\times 32$ lattice. Here it should be noted that the dimers at
the boundaries do not fluctuate much, and the central part of the system can be regarded as a kind of domain-wall state with de-facto plaquette
order. For system sizes larger than the domain wall thickness, the bonds in the center of the system fluctuate collectively between actual long-range 
ordered horizontal and vertical bond patterns. For very large lattices, the time scale of these fluctuations between the two different bond order 
realizations becomes too long (diverging exponentially with the system size) to observe in simulations, and the system may in practice become completely 
trapped in one of the sectors. This is seen in Fig.~\ref{Fig:BondStrength_SQ}(b) for an $80 \times 80$ system, where the central part of the system only 
exhibits strong $y$ dimers.

\begin{figure}[t]
  \includegraphics[width=70mm,clip]{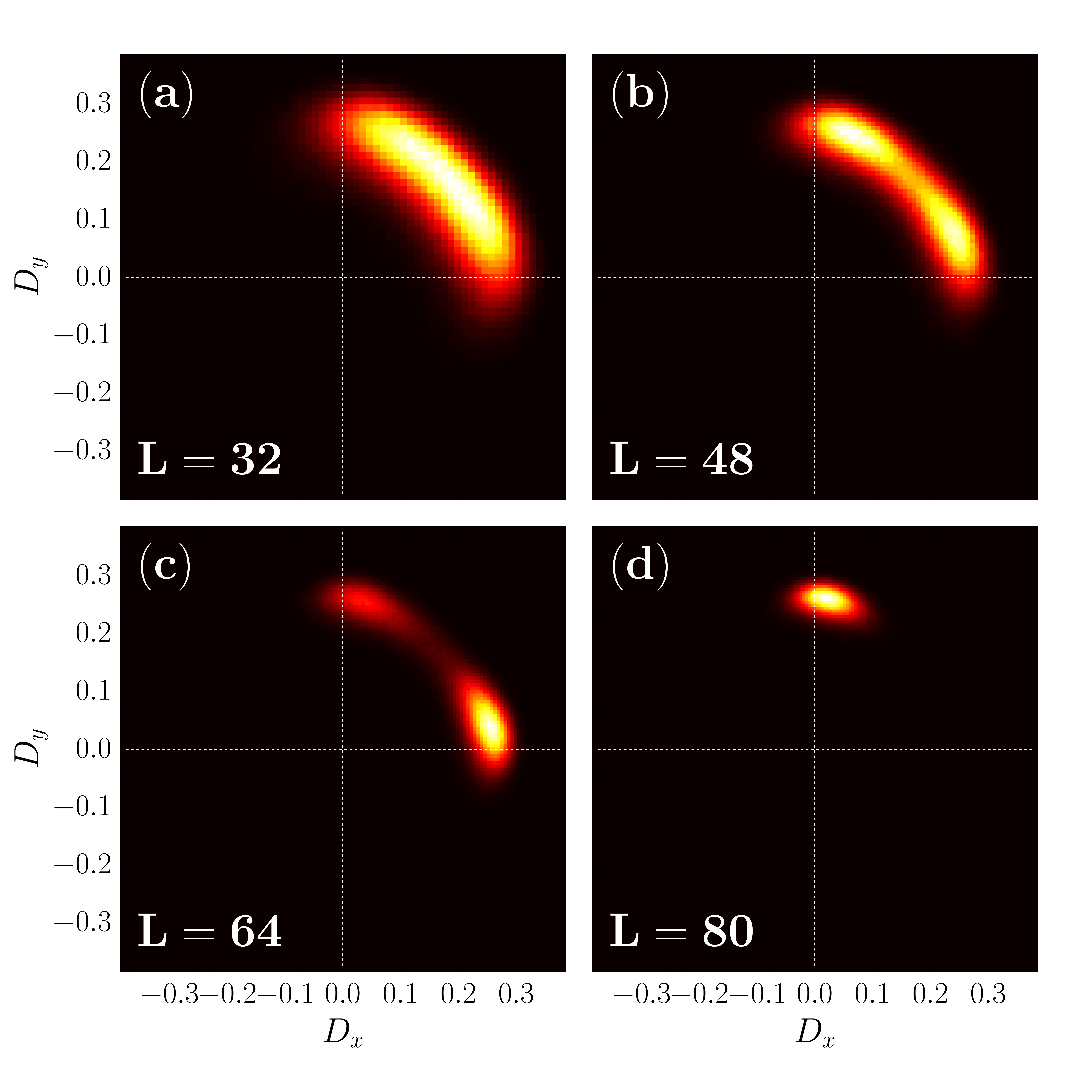}
  \vskip-2mm
  \caption{Distribution $P(D_x,D_y)$ with the order parameters $D_x$ and $D_y$ evaluated on the central 
  $L/2 \times L/2$ part of $L=32$ (a) and $L=80$ (b) lattices.}
  \label{Fig:Hist_SQ}
  \end{figure}

To further illustrate this symmetry breaking occurring in the simulations, 
in Fig.~\ref{Fig:Hist_SQ} we show the probability distribution $P(D_x, D_y)$ of the dimer order parameter as 
collected in the QMC process. For the smallest system, $L=32$, the peak in the distribution corresponds to equal $D_x$ and $D_y$, i.e., resonating plaquette 
order or equal amounts of static $x$ and $y$ dimers. For a slightly larger system, $L=48$, we observe the peak splitting into two, indicating a state that is
now fluctuating between $x$ and $y$ oriented bond order. The splitting between the peaks grows with increasing system size as the two patterns become
more dominated by the majority order, and the tunneling probability decreases (reflected in smaller weight close to the line $D_x=D_y$). For $L=64$, the 
two peaks have unequal density, due to the long time scale of fluctuation for this system size, and for the lagest system, $L=80$, the simulation was 
completely locked into the $y$ sector. As is typical in systems with a discrete symmetry of the order parameter, the time scale of tunneling between
sectors should grow exponentially with increasing $L$, and in practice it is not possible to sample equally both sectors for large systems.

\begin{figure}[t]
  \includegraphics[width=75mm,clip]{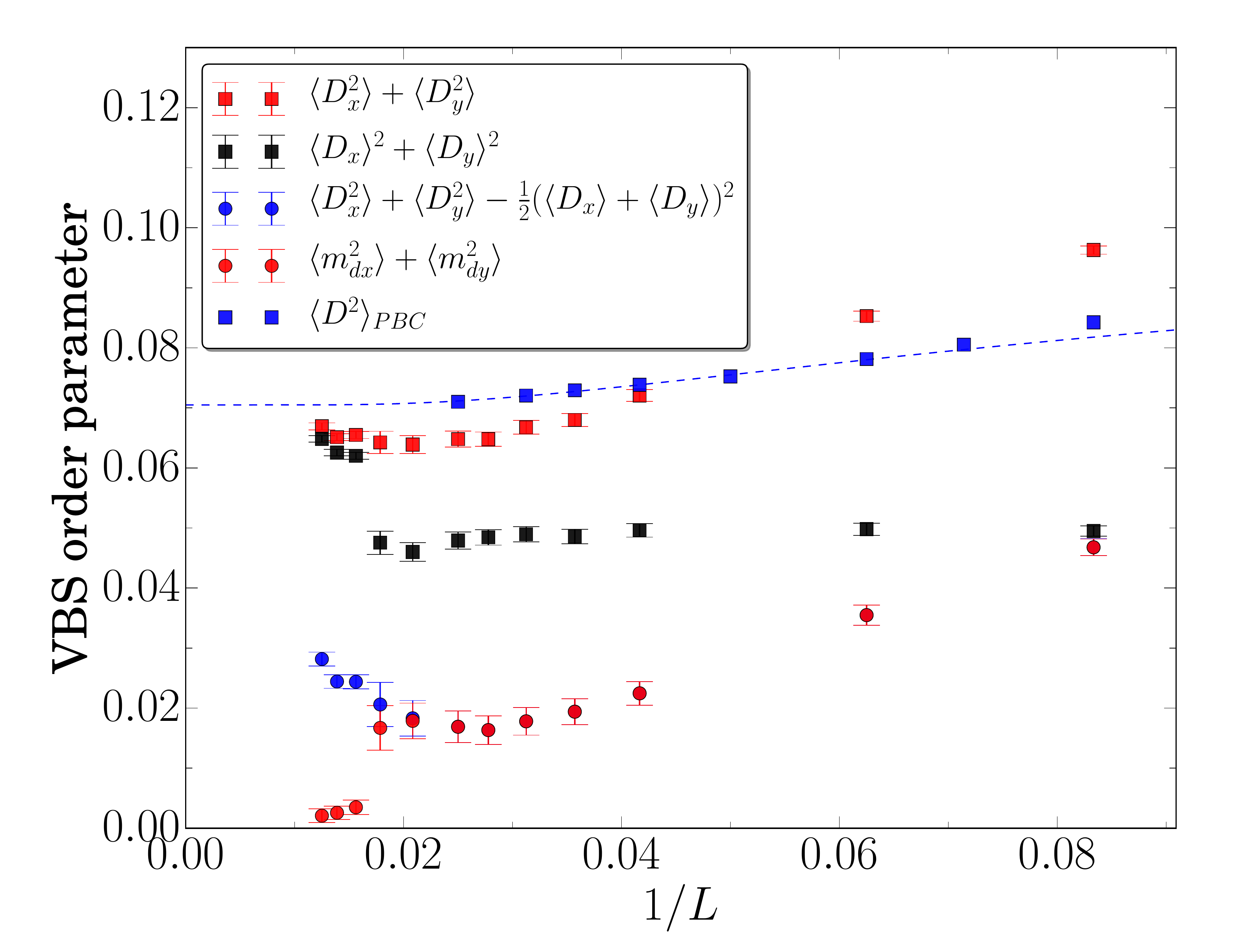}
  \vskip-2mm
  \caption{Dependence on the inverse system size of different definitions of the squared VBS order parameter. The quantities indicated
   by red and blue circles (the data sets with the smallest values) coincide almost exactly for system size up to $L=48$ and therefore only 
   the blue symbols are clearly visible.}
  \label{Fig:D_SQ}
  \vskip-5mm
\end{figure}

The method-related symmetry breaking is not a problem in practice, as long as computed quantities are insensitive to the symmetry breaking, e.g., with 
the definition $\langle D_x^2\rangle + \langle D_y^2\rangle$ of the VBS order parameter. As shown in Fig.~\ref{Fig:D_SQ}, results based on
this definition for the open system agrees with those for periodic boundary conditions in the limit $L \to \infty$, though the extrapolation to
infinite size is easier for the periodic systems. On the open systems the behavior is non-monotonic.
We also show results for $m_{dx}^2 + m_{dy}^2$, based on the definitions by Liu et al.~in 
Eq.~(\ref{vbsdef1}). In this case we see a sharp change in the behavior at a system size corresponding the the de-facto symmetry breaking of the QMC 
simulations for system sizes above $L=64$. For the smaller systems, the results appear to extrapolate to a non-zero value, but for the larger sizes 
the values drop rapidly toward zero. The latter behavior reflects the cancelation of the terms in the order-parameter definition, Eq.~(\ref{vbsdef1}), 
when the ground state is unique (in practice, due to the broken symmetry). For the smaller sizes the cancelation is not complete because of the two-fold 
degeneracy. A similar discontinuous behavior arising from symmetry breaking is seen in a symmetric definition based on the induced order parameter
(i.e., squaring the components after the mean value has been computed), $\langle D_x\rangle^2 + \langle D_y\rangle^2$, where
the results for the larger systems exhibit a jump up toward the results for periodic boundary conditions when the symmetry breaking takes place.
Before symmetry breaking we have $\langle D_x\rangle = \langle D_y\rangle$, and, because the squares are taken, a value $1/2$ of 
$\langle D_x^2\rangle + \langle D_y^2\rangle$ obtains (in the limit $L \to \infty$ without symmetry breaking).

A properly symmetrized version of the definition (\ref{vbsdef1}), in its equivalent form  (\ref{vbsdef2}), is 
$\langle D_x^2\rangle + \langle D_y^2\rangle - \frac{1}{2}(\langle D_x\rangle + \langle D_y\rangle^2$. In Fig.~\ref{Fig:D_SQ} it can
be seen that the results for this quantity concide with $m_{dx}^2 + m_{dy}^2$ when there is no symmetry breaking, while after symmetry 
breaking the two definitions diverge sharply. The properly symmetrized definition should be $3/4$ of the standard squared VBS order parameter 
for periodic boundaries in the thermodynamic limit.

\section{Conclusion}
\label{sec:conclusions}

We have discussed why the quantities $m_{dx}^2$ and $m_{dy}^2$ [Eq.~(\ref{vbsdef1})] used in Ref.~\cite{Liu18} 
may not capture long-range VBS order properly. It is clear that, in a system where the VBS order parameter symmetry is fully broken, the terms 
subtracted in Eq~(\ref{vbsdef1}) correspond to the actual order parameter of interest, and what is left vanishes for large distances (large systems).
In calculations with tesor network states, such as the PEPS used in Ref.~\cite{Liu18}, symmetry breaking can take place due to unequal treatment of 
the $x$ and $y$ directions or imperfect optimization (even on periodic lattices). As in Monte Carlo simulations, which may be trapped in one out of 
two or more sectors of the order parameter, this kind of ``artificial'' symmetry breaking may not be a problem in practice, as long as the consequences 
are understood and taken into account properly.

A VBS phase in the $J_1$-$J_2$ Heisenberg model cannot be excluded by the results 
presented in Ref.~\cite{Liu18}. Judging from other recent calculations with a variety of methods \cite{Gong14,Morit15,Wang18,Haghshenas18}, we expect VBS order 
in a narrow range of coupling ratios $g=J_2/J_1$ (roughly for $g \in 0.52,0.61$]). According to the same calculations, a gapless spin liquid may exist 
for slightly smaller values of $g$ (roughly for $g \in 0.45,0.52$]). It would be very interesting to see the VBS order parameter from the calculations
in Ref.~\cite{Liu18} without the subtraction of the crucial boundary induced contributions, as well as the boundary-induced order parameter itself.
We also point out that it may be advantageous to use rectangular lattices in PEPS calculations, as is evident from the behavior of $\langle D_y\rangle$ 
in Fig.~\ref{Fig:D}.

Aside from the use of a potentially flawed 
VBS order parameter, the calculations in Ref.~\cite{Liu18} are impressive and suggest that the method of Monte Carlo 
sampling of the physical degrees of freedom and gradient-based optimization \cite{Sandvik07,Schuch08,Wang11} may indeed be one of the most powerful 
ways to compute with tensor-netweok states. Very recently, further progress along these lines were reported in the context of the same $J_1$-$J_2$ 
Heisenberg model up to system size $24\times 24$ \cite{Liu19}. The VBS order was not discussed, however.

\null\vskip0mm
\centerline{\bf\small ACKNOWLEDGMENTS}
\null\vskip0mm

\noindent
We would like to thank Wenyuan Liu for discussions.
This work was supported by the NSF under Grant No.~DMR-1710170 and by a Simons Investigator Award.
The numerical calculations were carried out on Boston University's Shared Computing Cluster.

\end{document}